\documentclass[preprint,12pt,authoryear]{elsarticle}

\usepackage{amssymb}
\usepackage{tikz}
\usepackage{pgfplots}
\usepgfplotslibrary{statistics}
\usepackage{multirow}
\usepackage{eurosym}
\usepackage{float}
\usepackage{csquotes}
\usepackage{hyperref}
\usepackage[most]{tcolorbox}

\journal{Information and Software Technology}

\begin{document}

\begin{frontmatter}



\title{Innovation Initiatives in Large Software Companies: \\A Systematic Mapping Study \tnoteref{t1,t2}}
\tnotetext[t1]{This is the authors' version of the manuscript accepted for publication in the Journal of Information and Software Technology. Copyright owner's version can be accessed at \url{https://doi.org/10.1016/j.infsof.2017.12.007}. \textcopyright 2018. This manuscript version is made available under the CC-BY-NC-ND 4.0 license. \url{http://creativecommons.org/licenses/by-nc-nd/4.0}}
\tnotetext[t2]{Please cite as: Henry Edison, Xiaofeng Wang, Ronald Jabangwe, Pekka Abrahamsson (2018). Innovation Initiatives in Large Software Companies: A Systematic Mapping Study. \textit{Information and Software Technology}, 95:1--14. }


\author[unibz,ssrn]{Henry Edison\corref{cor1}}  \ead{henry.edison@inf.unibz.it}
\author[unibz,ssrn]{Xiaofeng Wang} \ead{xiaofeng.wang@unibz.it}
\author[usd,ssrn]{Ronald Jabangwe} \ead{rja@mmmi.sdu.dk}
\author[uoj,ssrn]{Pekka Abrahamsson} \ead{pekka.abrahamsson@jyu.fi}
\cortext[cor1]{Corresponding author}

\address[unibz]{Free University of Bozen-Bolzano, Piazza Domenicani 3, Bolzano 39100, Italy}
\address[usd]{Maersk Mc-Kinney Moller Institute, SDU Software Engineering, University of Southern Denmark, Campusvej 55, DK-5230 Odense M}
\address[uoj]{Faculty of Information Technology, P.O. Box 35, FI-40014, University of Jyv\"{a}skyl\"{a}, Finland}
\address[ssrn]{Software Startup Research Network \url{https://softwarestartups.org}}

\begin{abstract}



\textit{Context:} To keep the competitive advantage and adapt to changes in the market and technology, companies need to innovate in an organised, purposeful and systematic manner. However, due to their size and complexity, large companies tend to focus on the structure in maintaining their business, which can potentially lower their agility to innovate. 

\textit{Objective:} The aims of this study are to provide an overview of the current research on innovation initiatives and to identify the challenges of implementing those initiatives in the context of large software companies. 

\textit{Method:} The investigation was primarily performed using a systematic mapping approach of published literature on corporate innovation and entrepreneurship, which was then complemented with interviews with four experts with rich industry experience. 

\textit{Results:} Our mapping study results suggest that, there is a lack of high quality empirical studies on innovation initiative in the context of large software companies. A total of 7 studies are conducted in the context of large software companies, which reported 5 types of initiatives: intrapreneurship, bootlegging, internal venture, spin-off and crowdsourcing. Our study offers three contributions. First, this paper represents the map of existing literature on innovation initiatives inside large companies. The second contribution of this paper is to provide an innovation initiative tree. The third contribution is to identify key challenges faced by each initiative in large software companies. 

\textit{Conclusions:} At the strategic and tactical levels, there is no difference between large software companies and other companies. At the operational level, large software companies are highly influenced by the advancement of Internet technology. In addition, large software companies use open innovation paradigm as part of their innovation initiatives. We envision our future work is to further empirically evaluate the innovation initiative tree in large software companies. More practitioners from different companies should be involved in the future studies. 
\end{abstract}


\end{frontmatter}

\section{Introduction}
\label{sec:introduction}

How do large companies maintain their position in hyper competitive market? Over the years, corporate management relies on traditional way of advancement, which focuses on cost and lead time reduction and quality improvement \citep{rejeb08}. They are necessities but insufficient. Companies now operate in a time of increasingly tougher trading conditions, due to the expansion of the global market and technological advances \citep{kuratko15}. The advancement of Internet technologies has opened new markets worldwide and thus, increased competition among established companies \citep{thornberry01}.
 
Today, it is widely accepted that innovation is vital to companies to sustain their competitive advantages e.g. \citep{chandy00,teece07,kuratko14}. Innovation is ``the implementation of a new or significantly improved product (good or service), or process, a new marketing method, or a new organisational
method in business practices, workplace organisation or external relations'' \citep[p.46]{oecd05}. Being innovative allows companies not only to keep stable in the dynamic and disruptive environment but also to create new business opportunities.

Developing product innovation is a risky activity \citep{kleinschmidt91,song98,khurum15}. Many companies are too risk-averse to engage in any innovation initiatives \citep{ahmed98,gorschek10}. As in automated factories, people in large companies are trained to do prescribed and specific tasks reliably. Hence, any endeavour to change the status quo will encounter resistance. The implementation of an innovative idea must compete with other product development activities \citep{deVen86, connor13}.

The constraint of large companies to sustain innovation is not due to a lack of innovative ideas or employees \citep{pinchot85,menzel07}. In fact, large companies are considered the engine of innovation because of their ability to diversify and grow through internal development \citep{kacperczyk12}. Technology-based companies are limited by ``technology inertia'', which allows them to work only on the ideas within their scope \citep{ghemawat91}. Moreover, due to the size and complexity of the modern business, they tend to be bureaucratic, which can potentially lower a company's agility to innovate \citep{thornberry01}. The failure to generate innovation is also caused by ``the incumbent's curse'' \citep{chandy00}. Companies put their energy into products or technology, with which they are currently successful in the market so they are reluctant to take the risk to innovate by following different paths of innovation.

Given the plethora of literature on corporate innovation, we are interested in finding out what has been researched about innovation initiatives in large software companies. We define an innovation initiative as a risk-taking, proactive and innovative undertaking inside corporation \citep{covin91}. To achieve this goal, we used a systematic mapping study (SMS) and complemented it with the interviews with four experts with rich industry experience, to identify the common initiatives in large software companies performed to sustain innovation and the challenges in doing such initiatives in their context. The contributions of this study are three-fold. First, this study provides an overview of the current research on innovation initiatives inside large  software companies. The second contribution of this paper is to provide an innovation initiative tree. The third contribution is, our study also identifies key challenges faced by each innovation initiative in large software companies.

The remainder of this paper is structured as follows. Section \ref{sec:related_work} discusses the background and related work. The approach followed in the systematic mapping study and interviews is presented in Section \ref{sec:review_approach}. A summary of the mapping study is presented in Section \ref{sec:characteristics_primary_studies} and then followed by the results and analysis of data extracted from the mapping studies in Section \ref{sec:results}. The results from the interviews are reported in Section \ref{sec:initial_evaluation}. The findings are discussed in Section \ref{sec:discussion}. An outline of the conclusion and future work are presented in Section \ref{sec:conclusion}.
\section{Background and Related work}
\label{sec:related_work}

Today, large companies also must compete with new and emerging startups, which have become one of the key drivers of the economy and of innovation. In 2016, 550,000 new businesses or startups were established every month in the U.S. only \citep{fairlie16}. Even though they are inexperienced, young and immature \citep{sutton00}, their products are disrupting traditional markets and are putting well-established actors under pressure. Uber, Spotify, and Airbnb, to name just a few, are examples of software startups that have grown rapidly. Startups offer new products, new business models, and new business value at high speed, and with cutting edge technology. They continuously talk to their potential customers in order to discover gaps in the existing offers, iterate, and conduct experiments to find repeatable and scalable business models. They are willing to pivot immediately if the opportunity does not prove viable.

Together with operational excellence and customer intimacy, innovation is one of the three strategic options that companies need to prioritise when deciding on which unique value they want to bring to customers \citep{treacy95}. Through innovation, companies are able to create new markets and entry barriers, challenge market leaders and leapfrog competition \citep{brown95}. Innovation also allows companies to accrue high profit because at the time a new product is released, there is no competition in the market until imitators produce similar products \citep{roberts99}. 

The motivation of companies to initiate corporate innovation is the exceptional growth \citep{stevenson90}. Companies aggressively explore and exploit new opportunity even though they are limited by the availability of their resources. Thus, innovation must be done in an organised, purposeful and systematic manner \citep{drucker85}. It is not because of the work of a genius with brilliant ideas but a team with discipline to find the sources of innovative opportunities.

\cite{pavitt91} proposes four characteristics of large innovating companies. First is that large companies are a major source of technology and innovations. Their R\&D activities give big contributions in the design and operation of complex production technology. Second, that large companies have the capacity to recover from technological discontinuity. They have the ability to absorb and mobilise new skills and opportunities functionally and organisationally specialised to bring new innovations into the market. Third, to manage the complex business, large companies develop routines or standard procedure. Through their experience with changing world, they continuously learn to acquire new knowledge and competence. Forth, to get these new knowledge and competences, large companies allocate resources to pose unusual problems and difficult problems. 

Entrepreneurial activities within existing companies or corporate entrepreneurship (CE) have been suggested as a tool to facilitate companies' efforts to sustain innovation and improve their competitive positioning \citep{kuratko14}. CE is considered the process of creating new business (corporate venturing), strategic renewal or innovation within existing business \citep{sharma99}. Innovation is not necessarily required in either corporate venturing nor strategic renewal. For example, the idea for creating new business might come from imitating a successful product or service in the market. However, imitation does not lead to sustainable competitive advantage \citep{burgelman91}. This is contrary to what \cite{drucker85} contends that innovation is the source of entrepreneurship. Innovation makes the difference between entrepreneurial act and just opening new business. 

Like any modern dynamic business, software companies are highly influenced by its knowledge-intensive and technology-driven nature. For them, sustained innovation is critical, since innovation has become the main avenue for rapid growth \citep{nambisan02}. However, unlike in other high-technology domain, product life cycle in software industry is shorter, and often can be measured in months or weeks. Software is malleable and intangible and the threshold to enter the software market is low \citep{pikkarainen11}. Time is the main resource consumed to write, compile and test the code \citep{moe12}. Moreover, the boundary between service and product is imperceptible, which makes a significant impact of the potential evolutionary path for software companies \citep{nambisan02}.

Corporate innovation and CE are topics that still have received large attention until present. We identify several related works to our study. Prior to this work, multiple recent reviews on CE have been done e.g. \cite{corbet13,kuratko15}. In the software engineering domain, our previous work found several systematic literature reviews have been performed to aggregate different aspect of innovation \citep{edison16}. Through a systematic review, \cite{yague14} reported the existing assessment schemes applicable to software product innovation. In addition, \cite{munir15} presented a review on open innovation. We are different from these reviews by focusing specifically in the context of large software companies. We do not limit ourselves to a specific type of innovation e.g. product or process, but rather we look at broader innovation in large software companies.
\section{Review Approach}
\label{sec:review_approach}
The aims of this study are to provide an overview of the current research on innovation initiatives and to identify the challenges of implementing those types of initiative in the context of large software companies. To this end, Systematic Mapping Study (SMS) has been considered more appropriate and beneficial for this study than Systematic Literature Review (SLR) as the research approach. The purpose of a SMS is to investigate the literature on a field of particular interest for the purpose of determining the nature, scope and number of published primary studies \citep{petersen08}. It facilitates to obtain a broader view of wide and often poorly-defined research areas \citep{kitchenham07,petersen08}. Hence, the research questions for the mapping study are:
\begin{itemize}
	\item \textbf{RQ1:} \textit{What types and patterns of innovation initiative in the context of large companies exist in the literature?} The purpose for RQ1 is to get an overview of the current research on innovation initiatives. We broadened our scope to search for any type of large companies to ensure that we did not miss any relevant papers in the context of large software companies.
	\item \textbf{RQ2:} \textit{What are the challenges of implementing those types of initiative in the context of large companies?} RQ2 aims to identify the challenges of implementing those types of initiative in the context of large companies.
\end{itemize}

To complement the results of our SMS, interviews with experts from software industry were conducted to obtain more contextual understanding of different types of innovation initiatives in large software companies, and the challenges that companies encounter when implementing them.

\subsection{Search Strategy}
In this study, we used the following databases to perform the search:
\begin{itemize}
	\item Relevant to software engineering research: IEEEXplore, ScienceDirect, and Scopus. 
	\item Relevant to information system research: AIS eLibrary
\end{itemize}
	
To ensure that all the performed searches were consistent and comparable for each database, we used selected keywords and expressions derived from the research questions. Table \ref{table:search_strings} presents the generic combination of search strings to answer the research questions. The key term used in SMS is ``corporate innovation''. \cite{kuratko14} consider ``corporate innovation'' is the synonym of ``corporate entrepreneurship''. A study by \cite{sharma99} identifies different terms used in literature to describe corporate entrepreneurship, which includes intrapreneurship, internal corporate venture and spin-off.

The search was conducted in Feb 2016. We executed the search strings in different databases meeting their particular format requirements. All of the authors were involved in the identification of keywords and formulation of search strings.

\begin{table}[htbp]
    \caption{Search strings organisation}
    \label{table:search_strings}
    \begin{tabular}{|p{2cm}|p{10.8cm}|}
    \hline 
    Key term &  \texttt{``corporate innovation''} \\  \hline
    Synonym &  \texttt{``corporate entrepreneurship''} \\ \hline
    Narrower terms &  \texttt{ intrapreneur OR ``spin-off'' OR ``spin-out'' OR spinoff OR spinout  OR ``internal corporate venture'' OR ``open innovation''} \\ \hline
    Related terms &  \texttt{``corporate entrepreneur'' OR ``corporate entrepreneurial'' OR ``corporate innovation''} \\ 
    \hline 
    \end{tabular}
\end{table}

\subsection{Study Selection}
The potential primary studies were reviewed based on three phases. The first phase was intended to ensure the uniqueness of the article by removing duplicates, which was decided based on the similarity of title and authors. EndNote was used to identify the duplicates and complemented by a manual analysis. The first author was responsible for removing duplicates and non-research papers.

In the second phase, we reviewed the articles based on the title and abstract. The inclusion and exclusion criteria used in this phase are shown in Table \ref{table:criteria}. Each paper was evaluated by two reviewers. The first and second authors each reviewed the same paper. For a paper to be included the two reviewers had to be in agreement. In the cases where the reviewers did not agree on the inclusion or exclusion of the paper, a meeting was held with both reviewers present to discuss the appropriate action.

In the third phase, the remaining papers were reviewed in a similar manner as in the second phase by the first and the third authors. Papers with full-text and met the inclusion/exclusion criteria were assessed on their quality. Each paper was evaluated by two reviewers. The final set of papers obtained after this phase is the primary studies of this mapping study.

\begin{table*}[htbp]
    \caption{Inclusion and exclusion criteria}
    \center
    \label{table:criteria}
    \begin{tabular}{|p{6.3cm}|p{6.3cm}|}
    \hline 
    Inclusion criteria & Exclusion criteria \\ \hline
    $\bullet$ The articles describe, propose or evaluate a form of corporate innovation or corporate entrepreneurship. & $\bullet$ Studies on education or public sector and small and medium enterprises (SMEs) or national level. \\
    $\bullet$ Peer-reviewed papers, published in journal or conference & $\bullet$ Non research papers i.e. editorial, review, etc.  \\
    $\bullet$ Availability of full text written in English. & $\bullet$ Studies that discuss corporate innovation or corporate entrepreneurship in general and not specific to a particular type of innovation initiative. \\ 
    $\bullet$ Companies under study are considered as large companies & \\
       \hline 
    \end{tabular}
\end{table*}

We took the definition of large company as provided in each paper. However, to ensure that each paper was treated in the same way, we took the definition of large company as given in \cite{eu15}:  (1) staff headcount: employ $>$ 250 persons, and (2) annual turnover: $>$ \euro 50 million, or balance sheet total: $>$ \euro 43 million. Hence, for each paper we took the name of the company under the study, and looked into various sources including company website, newspaper or magazines to determine the information about its size in terms of number of employees or annual turnover.

\subsection{Quality Assessment}
The aim of the quality assessment is to assess the extent to which the selected primary studies have addressed bias and validity \citep{kitchenham07}. The quality assessment was used to include or exclude the papers. Each paper was assessed in the following aspects: research design, data collection, data analysis, context, results and conclusion. 

A checklist consisting of 9 questions covering the aforementioned aspects was developed to operationalise the assessment activity.  The checklist is based on the rigour of reporting (adapted from \cite{ivarsson10}): how well the reviewer is able to understand the research steps, details on how the corporate innovation or corporate entrepreneurship was performed in large software companies and the traceability of the research steps and the study findings and conclusion.  A list of questions for assessing the quality is captured in Table \ref{table:quality_assessment}.

\begin{table*}[htbp]
    \caption{Quality assessment criteria}
    \label{table:quality_assessment}
       \center
    \begin{tabular}{|p{4cm}|p{0.5cm}|p{7.5cm}|}
    \hline 
    Category & \multicolumn{2}{|c|}{Question} \\ \hline
    \multirow{2}{*}{Research Design} & 1 & Is the aim of the research sufficiently explained and well-motivated? \\ \cline{2-3}
	& 2 &    Is the research methodology clearly described? \\ \hline 
    Data  Collection &	3 & Are the data collection clearly described? \\ \hline 
    \multirow{3}{*}{Data Analysis} & 4 &   Is the data analysis used in the study adequately described? \\ \cline{2-3}
     & 	5a & Qualitative study: Are the interpretation of result clearly described? \\ \cline{2-3}
     &	5b & Quantitative study: Are the effect size reported with assessed statistical significance? \\ \hline
    Context &	6 & Is the type of corporate innovation or corporate entrepreneurship adequately described? \\ \hline
    \multirow{3}{*}{Results and Conclusions} & 7 & Are validity threats discussed? \\ \cline{2-3}
      & 8 & Does the empirical data and results support the conclusions?    \\ 
    \hline 
    \end{tabular}
\end{table*}

The quality of each paper was assessed by the first and third author. Each criterion was rated according to a ``Yes'' (indicating that data for the specific question is available), ``Somewhat'' (indicating that data is vaguely available) or ``No'' (indicating that data is unavailable), which corresponds to the score 2, 1 and 0. The quality assessment criteria distinguished between qualitative and quantitative study. However, the maximum score that a paper could get was 18. This would happen if the study employed both quantitative and qualitative approach.

A pilot with 5 papers was conducted before the actual assessment to check if both reviewers had the same understanding of each question. Any dissimilarity in assessment between reviewers was discussed until a consensus was reached.

\subsection{Data Extraction and Synthesis}
\label{sec:data_extraction}
In this study, the data is synthesised using narrative summary of the following aspects that were extracted from the selected primary studies:
\begin{itemize}
	\item Research type, classified as (modification of \cite{wieringa06}): 
		\begin{itemize}
			\item empirical research (combination of evaluation and validation papers);
			\item experience report (lesson learned by the author based on his/her experience);
			\item opinions (contains predominantly the author's opinion either as researcher or practitioners) and 
			\item conceptual framework (philosophical papers that describe new conceptual framework);
		\end{itemize}
	\item The context of the study, i.e. business domain, etc.;
	\item Type of corporate innovation or corporate entrepreneurship, including the name of the initiative, its description, the role of the initiators and the resources ownership;
	\item Challenges and benefits of using particular type of corporate innovation;
\end{itemize}

\subsection{Interviews}
Interview is a commonly used method in qualitative research to collect historical data from interviewee's memories, to gather the opinion or impression about something or to identify the terminology in a particular setting \citep{seaman99}. Interview can be conducted either by having face-to-face (one-on-one, in person) interview, telephone interview, focus group or email interview \citep{creswell09}. In this study, we conducted face-to-face interviews to grasp as much information we could get from the interviewees. We followed the protocol as described in \cite{creswell09} to conduct the interviews. Potential interview candidates were identified from the practitioners who attended the International Conference on Product-Focused Software Process Improvement (PROFES) 2015, in Bolzano, Italy, where the first two researchers reside. It presented a unique opportunity to us to access experts on software product innovation in large companies. Therefore, a convenient sampling strategy was used. Four industry practitioners in middle-managerial level eventually agreed on participating in the interview process, since they have long experience in undertaking innovation activities in their companies. During the interviews, the industry practitioners were asked to reflect back on their experience, what types of innovation initiatives have been done in their companies, and what the key challenges of those initiatives in their context are. The results achieved in the systematic mapping study were used to guide the interview process.  All interviewees have an extensive industrial experience, with more than 10 years of experience in large software companies. Each interview lasted approximately 30-45 minutes. All interviews were transcribed verbatim. The profiles of the interviewees are shown in Table \ref{tab:interviewees_information}. 

\begin{table*}[htbp]
    \caption{Background information of interviewees}
    \label{tab:interviewees_information}
           \center
    \begin{tabular}{|p{2cm}|p{4cm}|p{6cm}|}
    \hline 
    Interview ID & Experience & Role and responsibility\\
    \hline 
    INT\_1 & $>$20 years of industrial experience & Director, leading a product development division \\ \hline
    INT\_2 & $>$10 years of industrial experience & Product line manager and architect \\ \hline
    INT\_3 & $>$25 years of industrial experience, $>$ 2 years of academic experience & Researcher, involved in various positions in the company, from technical staff to management position. Currently working as researcher in university. \\ \hline
    INT\_4 & $>$10 years of industrial experience & Director, heading innovation department \\ 
   \hline 
   \end{tabular} 
\end{table*}

\subsection{Validity Threats}
Our study is not impervious to threats to validity, which may affect the outcome of this study. In the following section, the threats to validity of this study will be identified and discussed.
 
\subsubsection{Selection Bias}
To mitigate selection bias, we tested various versions of search strings. Since innovation is a very broad term, we omitted the keyword ``innovation initiative''. From our pilot study, we found that the result was very abstract, generic and broad thus we often found a large set of search results consisting of mostly irrelevant papers. We did not preselect the journals and relied on the journals included in the five databases used. We also did not include any terms related to large software companies into the search string since the result was very small and most of them were not relevant to this study. This information was retrieved during data extraction phase.

\subsubsection{Reviewer Bias}
To limit subjective bias from an individual reviewer, each paper was reviewed by two reviewers when applying inclusion and exclusion criteria. In addition, prior to actual selection of primary studies, the first three reviewers performed pilot runs. The aim was to ensure that each reviewer had the same interpretation of the criteria to include or exclude the collected studies.

Two pilots were conducted before applying inclusion and exclusion criteria. The first pilot was in the second phase when applying inclusion and exclusion criteria on the title and abstract. The pilot was done with 20 papers and the reviewers held meeting after each pilot to discuss the experience of applying the criteria and resolved disagreement on how to interpret the criteria. Kappa value was computed to measure the level of agreement \citep{landis77}. The Kappa value for the first pilot showed that there was a fair agreement between the reviewers because there was differences in interpreting the criteria. Both reviewers discussed together and agreed to redo the pilot study. As a result, there was a good agreement level between the reviewers. 

The second pilot was in the third phase between when conducting full-text review. The reviewers selected five papers and applied the inclusion and exclusion criteria, together with the quality assessment. Only one pilot was performed because after the discussion, the reviewers knew that they were in a good agreement in the interpretation of both inclusion and exclusion and quality assessment criteria. 

\subsubsection{Reliability of Findings}
While conducting this type of study, there is a possibility of missing relevant papers. Among other reasons, researchers may use different terminology for a particular topic \citep{wohlin14}. However, we implemented various measures to mitigate this issue. ScienceDirect and Scopus are two databases that contain studies from various research areas, including software engineering and computer science. Therefore, during the search process, we did not restrict the mapping study to software related research area only, but also include literature from other important fields i.e. management, business and economics. 

Since this study involves many fields and hence a large number of potentially irrelevant studies, we defined inclusion /exclusion and relevance criteria, which was applied on the results of the search strings. The criteria were formulated explicitly and as clear as possible to avoid misunderstanding. The criteria was reviewed and evaluated by the second author to check whether they were too strict or too loose. Based on these criteria, the first author selected the relevant studies individually. We piloted the selection criteria to see whether there was an agreement among the authors and also whether the search strings had covered the main area we aimed for. A defined data extraction strategy was also used after piloting, in order to conduct a structured data extraction.

\subsubsection{Selection of Interview Participants}
All interviewees had extensive knowledge about innovation initiative in their companies and even involved in particular initiatives. However, our interview is vulnerable to internal validity threat, which is the selection of participants. We did not have access to more experts in the period of this study. 

We acknowledge that the number of interviewees is a limitation of this study. Nonetheless, the vast years of experience that the interviewees have (i.e., two over 20 years and two over 10 years) and the systematic approach for reviewing literature that we followed, which we provided detailed in order to enable replicability, should reduce the threat to validity. However, as part of future plan, more interviews are required in order to improve the generalisability of the findings.
\section{Characteristics of Primary Studies}
\label{sec:characteristics_primary_studies}

\subsection{Search Results}
Table \ref{table:search_result} shows the search string used in each digital library and the corresponding search results. The first author was responsible for performing the database search. We retrieved 4705 articles from all databases. 

\begin{table*}[htbp]
    \caption{Search results}
    \label{table:search_result}
    \center
    \begin{tabular}{|p{2cm}|p{9cm}|p{1.5cm}|}
    \hline
    Database & Search string & Total articles found\\ \hline
    IEEE &  \texttt{``corporate entrepreneur'' OR ``corporate entrepreneurship'' OR ``corporate entrepreneurial'' OR intrapreneur* OR ``spin-off'' OR ``spin-out'' OR spinoff* OR spinout* OR ``open innovation'' OR ``internal corporate venture''} &    585 \\ \hline
     ScienceDi-rect &  \texttt{tak(``corporate entrepreneur'' OR ``corporate entrepreneurship'' OR ``corporate entrepreneurial'' OR intrapreneur OR ``spin-off'' OR ``spin-out'' OR spinoff OR spinout OR ``open innovation'' OR ``internal corporate venture'')} & 1714 \\ \hline
    Scopus &  \texttt{TITLE-ABS-KEY(``corporate entrepreneur'' OR ``corporate entrepreneurship'' OR ``corporate entrepreneurial'' OR intrapreneur OR ``spin-off'' OR ``spin-out'' OR spinoff OR spinout OR ``open innovation'' OR ``internal corporate venture'')} &  2148 \\ \hline
    AIS e-library &  \texttt{``corporate entrepreneur'' OR ``corporate entrepreneurship'' OR ``corporate entrepreneurial'' OR intrapreneur OR ``spin-off'' OR ``spin-out'' OR spinoff OR spinout OR ``open innovation'' OR ``internal corporate venture''} & 258 \\ \hline 
    \end{tabular}
\end{table*}

\begin{figure*}[htbp]
    \centering
    \includegraphics[width=\textwidth]{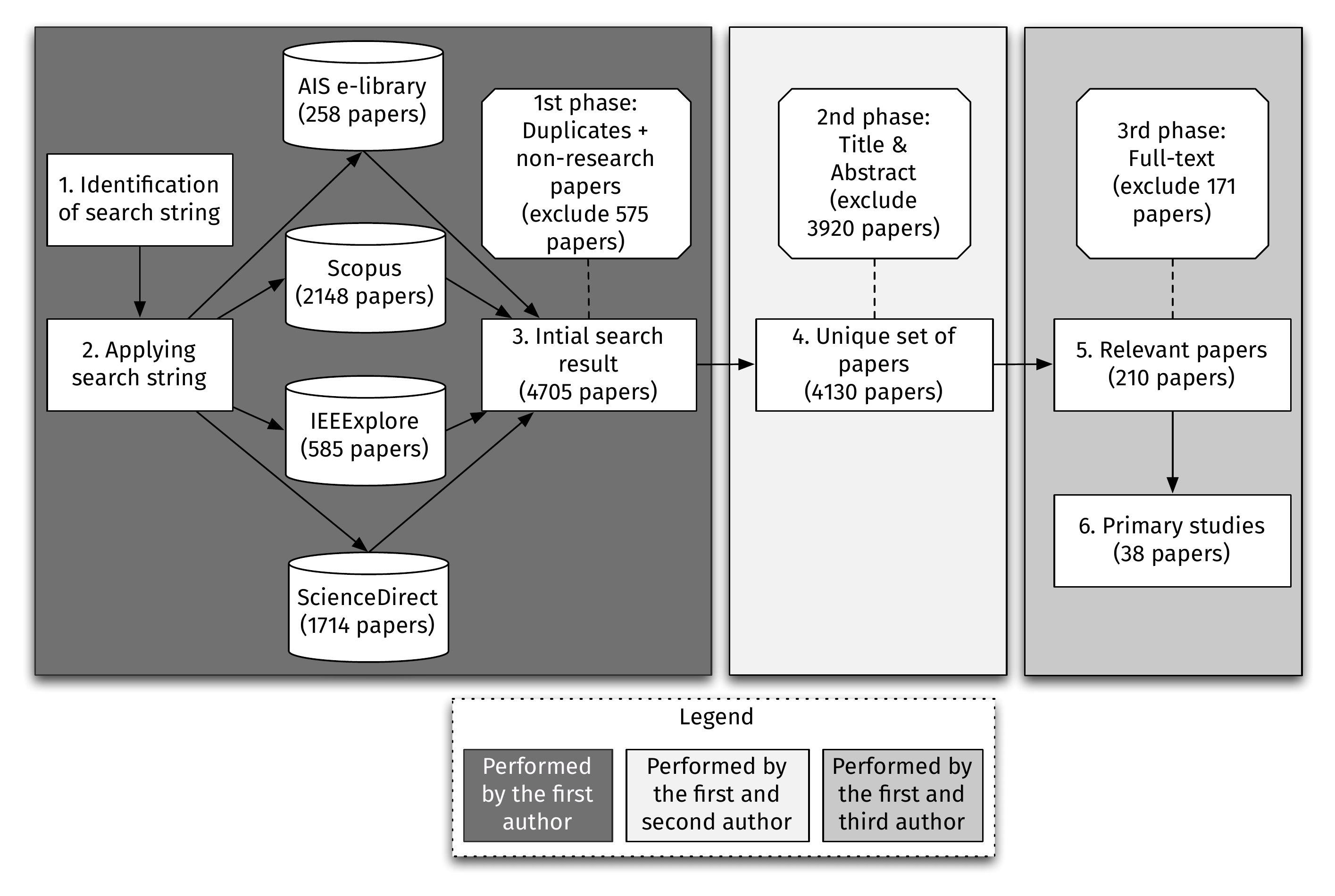}
    \caption{Selection process of primary studies}
    \label{fig:selection_procedure}
\end{figure*}

In the first phase, the first author removed 575 duplicates and non-research papers. The first and the second authors were responsible in evaluating the remaining papers based in the tile and abstract. The third phase was conducted by the first and the third authors. The selection process of the primary studies is shown in Fig. \ref{fig:selection_procedure}. By applying the inclusion/exclusion criteria sequentially, 38 papers were accepted as primary studies.  The 38 primary studies are listed in Table \ref{table:primary_studies} and referred using their IDs throughout the rest of the paper. Studies that are specifically conducted in large software companies context are written in bold.

\begin{table*}[htbp]
\centering
    \caption{List of primary studies}
    \label{table:primary_studies}
    \begin{tabular}{|p{0.7cm}|p{5.3cm}|p{0.7cm}|p{5.3cm}|}
    \hline 
    ID & Author(s) & ID & Author(s) \\ \hline 
    S1 & \cite{kierulff79} & \textbf{S20} & \textbf{\cite{huyskens07}}\\ \hline
    \textbf{S2} & \textbf{\cite{pinchot85}} & S21 & \cite{menzel07} \\  \hline
    S3 & \cite{ross87} & S22 & \cite{frederiksen08} \\  \hline
    S4 & \cite{duncan88} & \textbf{S23} & \textbf{\cite{loebbecke08}} \\  \hline
    S5 & \cite{mcgrath95} & S24 & \cite{buenstorf09} \\  \hline
    S6 & \cite{badguerahanian95} & S25 & \cite{kuratko09} \\ \hline
    S7 & \cite{birkinshaw97} & S26 & \cite{ford10} \\  \hline
    \textbf{S8} & \textbf{\cite{abetti97}} & S27 & \cite{anokhin11a} \\  \hline
    S9 & \cite{birkinshaw98} & S28 & \cite{anokhin11b} \\ \hline
    S10 & \cite{antoncic01} & S29 & \cite{clarysse11} \\ \hline
    S11 & \cite{abetti02} & S30 & \cite{espinosa11} \\ \hline
    S12 & \cite{chasteen03} & S31 & \cite{hasegawa11}\\ \hline
    \textbf{S13} &  \textbf{\cite{chesbrough03}} & \textbf{S32} & \textbf{\cite{jung11}} \\ \hline
    S14 &\cite{parhankangas03} & S33 & \cite{andersson12} \\ \hline
    S15 & \cite{mukherjee04} & S34 & \cite{berchicci13} \\ \hline
    S16 & \cite{augsdorfer05} & S35 & \cite{cazares13} \\ \hline
    S17 & \cite{christensen05} & S36 & \cite{digout13} \\ \hline
    S18 & \cite{subramanian05} & S37 & \cite{amundsen14} \\ \hline
   \textbf{S19} & \textbf{\cite{dahlander06}} & S38 & \cite{knoskova15} \\ 
    \hline 
    \end{tabular}
\end{table*}

\subsection{Publication Venues and Year}
The distribution of 38 primary studies is shown in Fig. \ref{fig:venues_year}.  According to our mapping study the earliest paper on corporate innovation is by \cite{kierulff79}. The paper recognises the role of corporate entrepreneur in leading a company into a significant new area of market and product. The figure also shows that there has been some interest on innovation and entrepreneurship for more than four decades and there are indications that more related publications will appear in the future. Most of the studies are published in a journal (82.5 \%) rather than conference (17.5\%).

\pgfplotsset{every tick label/.append style={font=\scriptsize}}
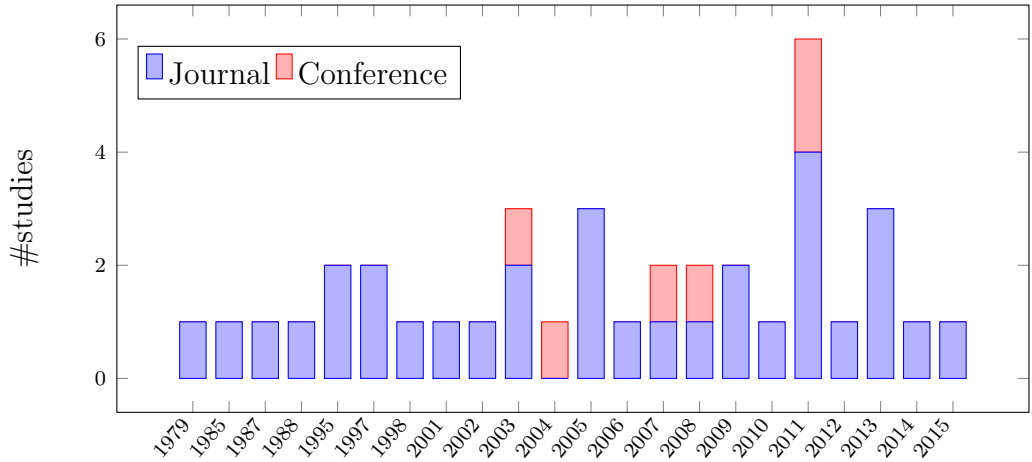
\begin{figure*}[htbp]
    \centering
\begin{tikzpicture}
\begin{axis}[
width=\textwidth,
height=7cm,
    ybar stacked,
    legend style={at={(0.2,0.90)},
      anchor=north,legend columns=-1},
    ylabel={\#studies},
    symbolic x coords={1979,1985, 1987,1988, 
		1995, 1997,1998, 2001, 2002, 2003, 2004, 2005, 2006, 2007, 2008, 
		2009, 2010, 2011, 2012, 2013, 2014, 2015},
    xtick=data,
    x tick label style={rotate=50,anchor=east},
    ]
\addplot+[ybar] plot coordinates {(1979,1) (1985,1) 
  (1987,1) (1988,1) (1995,2) (1997,2) (1998,1) (2001,1) (2002,1) (2003,2) (2004,0)
  (2005,3) (2006,1) (2007,1) (2008,1) (2009,2) (2010,1) (2011,4) (2012, 1) (2013,3) (2014,1) (2015,1)};
\addplot+[ybar] plot coordinates {(1979,0) (1985,0) 
  (1987,0) (1988,0) (1995,0) (1997,0) (1998,0) (2001,0) (2002,0) (2003,1) (2004,1)
  (2005,0) (2006,0) (2007,1) (2008,1) (2009,0) (2010,0) (2011,2) (2012, 0) (2013,0) (2014,0) (2015,0)};
\legend{\strut Journal, \strut Conference}
\end{axis}
\end{tikzpicture}
   \caption{Publication venues and year}
    \label{fig:venues_year}
\end{figure*}

\subsection{Quality of the Primary Studies}
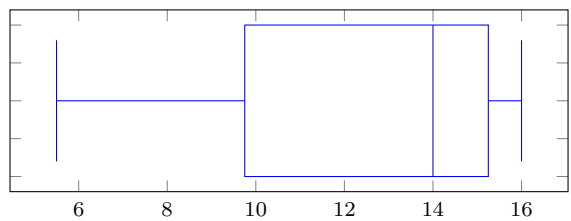
\begin{figure}[htbp]
	\centering
	\begin{tikzpicture}
        \begin{axis}
            [
            height=4cm,
            width=9cm,
            yticklabels={0},
            ]
            \addplot+[
            boxplot prepared={
              median=14,
              upper quartile=15.25,
              lower quartile=9.75,
              upper whisker=16,
              lower whisker=5.5
            },
            ] coordinates {};
            
          \end{axis}
    \end{tikzpicture}
	\caption{Percentile rankings of the quality scores}
	\label{fig:box_plot}
\end{figure}

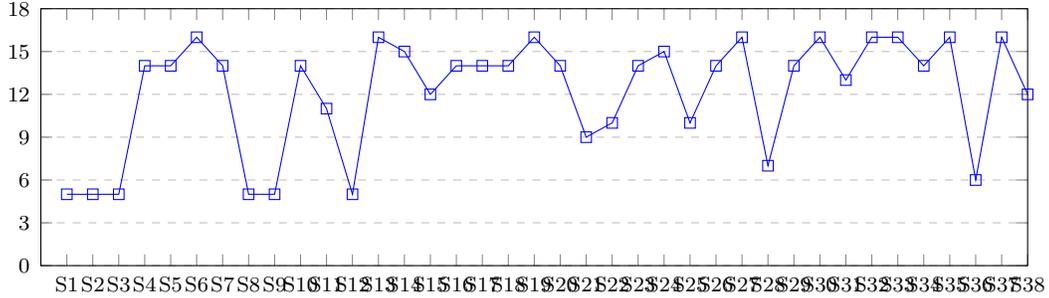
\begin{figure*}[htbp]
	\begin{tikzpicture}
		\begin{axis}[
			height=5cm,
            		width=14.7cm,
    			xmin=0, xmax=76,
    			ymin=0, ymax=18,
xtick={0,2,4,6,8,10,12,14,16,18,20,22,24,26,28,30,32,34,36,38,40,42,44,46,48,50,52,54,56,58,60,62,64,66,68,70,72,74,76},
    			ytick={0,3,6,9,12,15,18},  xticklabels={,S1,S2,S3,S4,S5,S6,S7,S8,S9,S10,S11,S12,S13,S14,S15,S16,S17,S18,S19,S20,S21,S22,S23,S24,S25,S26,S27,S28,S29,S30,S31,S32,S33,S34,S35,S36,S37,S38},
    			legend pos=north west,
    			ymajorgrids=true,
    			grid style=dashed,
		]
		\addplot[
    			color=blue,
   	 		mark=square,
    		] coordinates {
    			(2,5)(4,5)(6,5)(8,14)(10,14)(12,16)(14,14)(16,5)(18,5)(20,14)(22,11)(24,5)(26,16)(28,15)(30,12)(32,14)(34,14)(36,14)(38,16)(40,14)(42,9)(44,10)(46,14)(48,15)(50,10)(52,14)(54,16)(56,7)(58,14)(60,16)(62,13)(64,16)(66,16)(68,14)(70,16)(72,6)(74,16)(76,12)
    };
\end{axis}
\end{tikzpicture}
\caption{Distribution of quality scores of all primary studies}
	\label{fig:quality_scores}
\end{figure*}

In general, based on the quality scores, the primary studies can be considered to be of good quality. The percentile rankings of the quality scores are shown in Fig. \ref{fig:box_plot}. No study got the maximum score (18), since none of the study was using both quantitative and qualitative approach. Studies with scores below the lower quartile lacked clear information about the research design, data collection and data analysis as required by Q1-Q5b of quality assessment criteria. These types of study are considered as opinions as described in Section \ref{sec:data_extraction}. Moreover, most of the studies within the inter-quartile range did not discuss validity threats and how they are mitigated, which negatively affected the trustworthiness of the reported findings \citep{robson11}. The distribution of quality scores of all primary studies is shown in Fig. \ref{fig:quality_scores}.

\subsection{Innovation Type and Context and Research Type}
Out of 38 primary studies, we extracted 9 different types of innovation initiative in large companies, together with the context where the studies take place. The types of innovation initiative were identified during the data extraction process.The names of these initiatives types were taken as explicitly described in the reviewed papers. As shown in Table \ref{table:quality_assessment}, question 6 (Context) ensured that the primary studies discussed at least one type of innovation initiative. This information was later extracted from the primary studies and used in the data synthesis. The classification of primary studies based on innovation types and study context are presented in Table \ref{tab:innovation_context}. We found that some studies are based on researchers' opinions and do not explicitly mention the domain where the initiatives take place. Hence, we identified with ``not specified''. On the other hand, some studies discussed innovation initiative in various types of industry, for example manufacturing, electrical and computer. Therefore, we categorised them into ``multiple industry''. 

\begin{table*}[htb]
\small
    \caption{Innovation type and context}
    \label{tab:innovation_context}
    \begin{tabular}{|p{1.8cm}|p{1.3cm}|p{1.3cm}|p{1.4cm}|p{1.1cm}|p{0.9cm}|p{1.4cm}|p{1.1cm}|}
    \hline
    \multirow{2}{*}{\parbox{1.8cm}{Innovation Type}} & \multicolumn{7}{c|}{Context} \\ \cline{2-8}
    & Software  & Enginee-ring & Manufac-turing & Energy & Finan-ce & Multiple Industry & Not Specified \\ \hline
    Intrapre-neurship & S2 & S18 & S15 & & & S4,S10, S37,S38 & S1,S3, S12,S21\\ \hline
    Bootlegging & S8 & & & & & S16 & \\ \hline
    R\&D & & S33,S34, S35 & & S22 & & & \\ \hline
    Internal Venture & S20,S23 & S6, S26 & & & S5 & S25 &\\ \hline
    Subsidiary & & & & & & & S7, S9 \\ \hline
    Joint venture & & & & & S30 & & \\ \hline
    Venture Capital & & S31 & & S29 & & & \\ \hline
    Spin-Off & S13 & S11,S17, S24 & & & &  S14,S27, S28 &  \\ \hline
    Crowdsour-cing & S19,S32 & & & & & S36 & \\ 
      \hline
    \end{tabular}
\end{table*}

Table \ref{tab:innovation_research} presents a classification of primary studies based on the innovation type and research type as described in Section \ref{sec:data_extraction}. More than 70\% of the primary studies (28 studies) are empirical research. 

\begin{table*}[htb]
    \caption{Innovation type and Research type}
    \label{tab:innovation_research}
    \center
    \begin{tabular}{|p{3cm}|p{2.1cm}|p{2cm}|p{2cm}|p{2.2cm}|}
    \hline 
     \multirow{2}{*}{Innovation Type} & \multicolumn{4}{c|}{Research Type} \\ \cline{2-5}
     & Empirical Research & Experience Report & Opinion & Conceptual Framework \\ \hline
    Intrapreneurship & S10, S15, S18, S37, S38 & S2 & S1, S3, S4, S12, S21& \\ \hline
    Bootlegging & S8, S16 &  & & \\ \hline
    R\&D & S22, S33, S34, S35 &  &  & \\ \hline
    Internal Venture & S5, S6, S20, S23, S25, S26 & & & \\ \hline
    Subsidiary & & S7, S9 & & \\ \hline
    Joint venture & S30 & & & \\ \hline
    Venture Capital & S29, S31 & & &  \\ \hline
    Spin-Off & S11, S13, S14, S17, S24, S28 & & & S27 \\ \hline
    Crowdsourcing & S19, S32 & & S36 & \\ \hline
    \end{tabular}
\end{table*}
\section{Results and Analysis}
\label{sec:results}

\begin{figure*}[htbp]
    \centering
    \includegraphics[width=\textwidth]{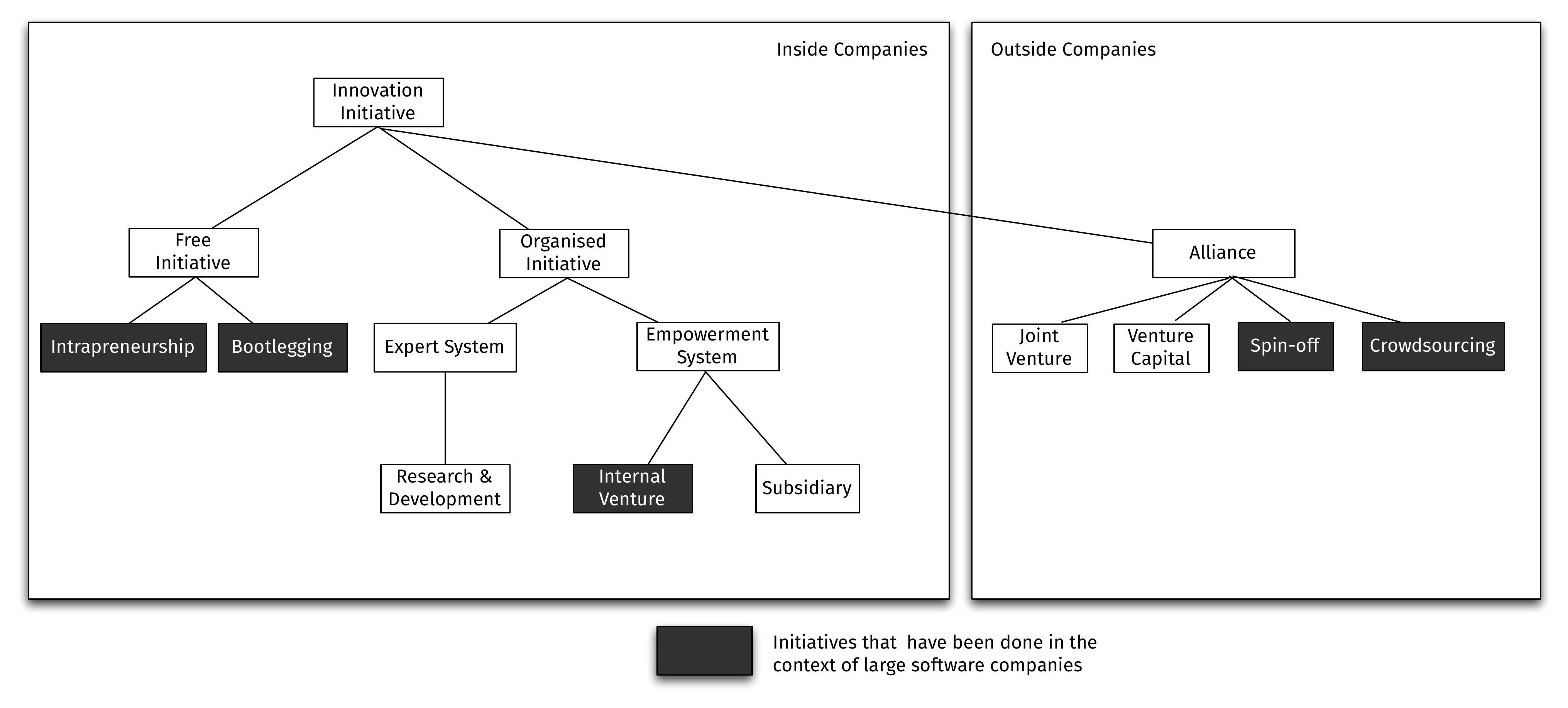}
    \caption{Innovation initiative tree}
    \label{fig:tree}
\end{figure*}

The reviewed studies allow us to draw an innovation initiative tree in large companies, which is presented in Fig. \ref{fig:tree}. The 9 innovation initiatives are represented as the leaves. Among the 9 innovation initiatives, 5 of them are found in the context of large software companies: intrapreneurship, bootlegging, internal venture, spin-off and crowdsourcing. 

As described in Section \ref{sec:data_extraction}, while looking into each of these initiatives, we also extracted the role of the initiators and the resources ownership. The role of the initiators refers to the role of ones who take the responsibility to start the initiative, e.g. the employees or the corporate management. The resources ownership defines who owns the resources used in the initiatives. A resource is one of the sources of competitive advantages \citep{barney91}. It is referred as all the assets (tangible and intangible) owned by a company i.e. brand names, employee, capital, in-house knowledge or technology etc. \citep{wernerfelt84}. 

By considering the role of initiator and the resource ownership, the 9 initiatives are further organised into a tree structure. Based on the resources ownership, innovation initiative can be done inside or outside the company. When companies have all resources needed, they tend to develop in-house. When companies lack resources to innovate, a strategic alliance might be a solution \citep{teng07}. Through alliance, companies get access to the required sources provided by other partners. In addition, each partner can learn how to develop new resources and competence needed to innovate \citep{gulati00}.

Based on the role of the initiators, inside company, innovation initiatives fall into free and organised initiatives. Free initiative is bottom-up approach and consisting of a series of un-organised activities. Employees come up with their own innovative ideas and try to convince management for approval \citep{sundbo96}. When the ideas are rejected by management, employee has choice to continue working in bootlegging activities, leave company or create a spin-off. If the idea is approved by management, employee initiates the development activities inside the company. 

In organised initiative, top management is responsible to nurturing and fostering internal innovation initiative. There are two types of organised initiatives: expert system and empowerment system. In expert system, innovation is usually carried out by a specific and dedicated unit inside the company who is responsible for developing new products, i.e. Research and Development (R\&D). On the other hand, in the empowerment system, innovation is generated through different types of initiatives: internal venture or subsidiary. 

\subsection{Innovation Initiative Types}
\label{sec:innovation_type}
The following sections discusses the current innovation initiative types in large companies.

\subsubsection{Intrapreneurship}
\label{sec:intrapreneurship}
Large companies do not lack innovative ideas, but they are poor in turning the ideas into new products (S2). In many cases, implementation of new ideas often bogs down in bureaucracy where analysis and approvals become mandatory. One way to address this issue is through intrapreneurship (S1,S2,S3,S4,S10,S12,S15,S18,S21,S37,S38). Intrapreneurs have the vision for new products and act on their vision as if they had their own companies: build the development team and run the business (S2). Moreover, intrapreneurs also push for change and develop creative responses in the company (S21). They are differ from innovation managers. While innovative managers look for new products for existing markets or new markets for existing products, intrapreneurs look for new products for new markets (S1).

Although the importance of intrapreneurs is recognised, companies struggle to nurture them. There are two main problems at strategic and tactical level (S1). At strategic level, management must recognise that creative employees may work in unpredictable ways. Some are great visionaries and willing to pursue them but some are very effective to imitate an idea and adapt it to a new setting. Some are very creative to seek a gap in the current market and fill it \citep{myers84}. Hence, management must support, facilitate and encourage entrepreneurial behaviour \citep{kuratko14}.

At tactical level, companies do not have proper ideas to reward creative employees. While most employees are looking forward to higher position (which reflects to higher responsibilities and rewards), intrapreneurs are more interested in freedom or autonomy on time and style how to accomplish their work (S4). Usually, employees are considered failing if they do something wrong in their work. Moreover, they do not get penalties even though they loose an opportunity. This policy does not support the creation of entrepreneurial spirit in the company. Failure must be considered as part of learning process. Otherwise, intrapreneur does not have incentive to innovate (S2).

Intrapreneurs can be anyone at any level and function in the company, who behave with entrepreneurial spirit (S21). They must be motivated which can be done by trust caring. Employees are becoming more motivated to undertake entrepreneurial activities when managers display confidence and satisfaction about the entrepreneurial projects and vice versa \citep{brundin08}.

\subsubsection{Bootlegging}
Bootlegging (or underground or skunkworks (S8)) refers to the innovation activity that is hidden from management until its introduction \citep{knight67}. The objectives of bootlegging are pre-research, product and process improvement, troubleshooting, new product and process development and purely scientific research (S16). In some cases, bootlegging is encouraged and promoted by management to overcome the bureaucracy and inertia against change (S8).  In other cases, bootlegging activities are initiated when an idea is rejected by top management (S8). Rather than quit and leave the company, the employee decides to startup the development underground.

Bootlegging activities need a champion to secure the resource procurement (S8). A champion is referred as the person who protects them from any interference in the company e.g. top executive. A champion can be anyone in the company. However, champion from higher management is more effective to ensure the sustainability of innovation process in the company.

In some cases, bootlegging is carried out without approval from management. As the consequence, most of these projects have limited access to existing resources (S16). In addition, higher management are needed to be assured that new ideas are related to the company knowledge. The level of uncertainty is similar as regular R\&D-based innovation (S16).

\subsubsection{Research \& Development (R\&D)}
In large and high-tech industry, innovative activities are performed by a specialised and dedicated entity, typically R\&D department. In R\&D, most innovations are scientific and/or technological based. The involvement of companies in R\&D activities are driven by the need to improve current process or products, researching new process or technology or specific user need. In fact, economies of scale in R\&D, risk diversification and access to greater financial success are the main benefits of large companies to generate radical innovation (S26). Unlike in high-tech industry, low-tech companies prefer to buy the technology rather than involve in R\&D, since their competition is about marketing not in the technology itself (S35).

When the technology becomes more advanced and complex, R\&D are demanded to bring more innovative products. However, not all technologies produced by R\&D are inline with and directly support the business goal. These technologies are called misfit technologies (S28). When this happens, the company has three options: keep scientific research, sell the technologies outside or introduce spin-off (S11,S28).

The study by \cite{gassmann10} identifies a trend that the complexity of technology has emerged needs of an alliance and partnership in conducting R\&D. Alliances are not only for cost saving but also for value creation. While being involved in an alliance, a company has two options: proactive attention by actively submitting suggestions or reactive attention by listening to external suggestions \citep{dahlander14}. By submitting suggestions together with valuable information, a company opens a channel to communicate with alliances to share, evaluate and develop the ideas. In reactive attention, the suggestions are used to complement the existing knowledge. However, the company must be able to balance between the time and effort needed to implement the ideas.

A particular mode of R\&D is an internal project. Project is defined as a temporary organisation of individuals to perform a complex, non-repetitive task and results in unique or highly customised output (S22). There are two types of internal projects: a base project, which is performed on the base of current market  or targeting at incremental innovation and base-moving project, to explore new market and exploit new technology or targeting at breakthrough innovation (S22). 

When a project is performed as the first in the purpose of diversification technology and market, it is called vanguard project (S22). It is not intended to improve company's operational excellency but as the testing mechanism and a learning process, by involving cross-functional team to generate new knowledge and capabilities. Vanguard project reduces the risk and uncertainty of entrepreneurship since it does not introduce new company and more focus on generating new knowledge. 

As the tool for radical innovation, vanguard project has two uncertainties: operational (internal factors e.g. team, features, etc) and environmental (external factors e.g. demand and technical change, knowledge transfers etc). Therefore, the traditional project management tools cannot be used in this project. The objective of vanguard project is to exploration not to achieve a set of predefined goals.

\subsubsection{Internal Venture}
\label{sec:internal_venture}
Internal venture refers to the introduction of new business within existing business to pursue product or market innovation (S25). The degree of newness is defined by new in the world and new in the industry (S25). New business can be established as the instrument to pursue incremental innovation (new product in current market or new market for current product) or radical innovation new product for new market).

The introduction of new venture is also seen as a core process to create new competence (S5). As a learning process, therefore failure is inevitable. Thus, the internal venture suggests to seek redirection rather than termination e.g. introduce new venture, since the new competence or knowledge might not be inline with main business stream.

The advancement of Internet technologies since 1990 has triggered the Internet entrepreneurship (S23). Companies like Bay, Google, Amazon, to name a few, emerge as e-venture which provided services via the Internet. The Internet entrepreneurship also emerges companies that offer community-driven service e.g. YouTube, MySpace, Facebook (S20,S23). Study S23 finds that the key enablers of e-venture are personal network, entrepreneurial team, business model, resources and marketing strategy. However, study S20 finds that e-ventures also deal with the same issue with classic internal ventures, i.e., support from top management and resources.

A particular mode of internal venture is incubator (S26).  An incubation provides an environment to generate novel ideas and to incubate them. Projects which show the potential disruptiveness and admits them to the incubator. In an incubation, activities are managed by a production team rather than a management team. When it comes to acceleration, companies can establish a new venture. 

Internal venture can be established through an acquisition of new or small companies with innovative products. For large companies, acquisition is the shortest way to bring new technology inside without having the need to develop it in-house. Acquiring independent company is also suggested by \cite{morse86} but with different motives. He argues that since intrapreneurship will not be success in large bureaucratic corporation, it is necessary to buy  independent startups, integrate them into the corporation and put intrapreneurs to grow. Hence, the intrapreneur has the autonomy to innovate, a place to do things differently without following the existing procedure. Moreover, having this separate division with its own resources will not generate more impact to the companies (S12).

Another reason to engage in acquisition is to have diversification. Acquisition may serve as a substitute for innovation \citep{hitt90}. It requires more resources causing less resources to invest in other initiatives. However, acquiring a new company is not trivial, instead it needs expertise and experience to decide on which companies to buy and their values. Otherwise, it might lead to performance declination \citep{hitt90}.

\subsubsection{Subsidiary}
In a large and geographical dispersal company setting, an innovation initiative might take place in the form of subsidiary initiative (S8).  The initiatives starts when the subsidiary companies identify a new business opportunity and sell it to the head office for a commitment to establish the business. There are two reasons for this initiative: market development and network optimisation. In market development, the new subsidiary are considered as the response to the need of local market. In network optimisation, the initiative aims to improve the current company global network internally. The improvement can be done through four ways:
\begin{itemize}
	\item Reconfiguration initiative: initiated by head office to support current market
	\item Maverick initiative: initiated without the approval from head office to support current market. For example, rather than selling only the own group's products, a subsidiary might sell competitors' product which is considered more competitive.  
	\item Bid initiative: creating new market, initiated jointly by head office and the subsidiary. The purpose is to be a leader in specific market.
	\item Leap of faith initiative: without approval from head office, try to identify new market and maintain internal market at the same time.
\end{itemize}

However, study S9 also identifies the cost of subsidiary initiative: empire building (rather than following the product lines decided by head office, subsidiary executives build different product which at the end affects globally its market positioning), lack of focus (too much entrepreneurship), cost of administering the internal market and internal unemployment.

\subsubsection{Joint venture}
While companies do not have the resources available to do entrepreneurship, joint venture (JV) can be the solution to obtain them. JV is a legal entity, established by two or more companies that share controls, profits and risks proportionally (S30). JV not only allows the company to obtain resources quickly, but also increase the capabilities needed to create new business. Through JV, a company can grow its size to strengthen its position in the market.

Study S30 finds that the main reason for companies to engage in JV are the entry to international markets. When the companies want to expand their market abroad, they need a cooperation with local partners. The local partners have more knowledge about local market, hence the cooperation is a short way of development. 

Although JV can be used to achieve wealth, S30 argues that it cannot be used as source of innovation. They find that the companies engaged in JV do not have intention to learn on the part of entrepreneurial form or the necessary knowledge to do particular things, for example opening new market or to put innovation into practice. When it comes to innovation, most companies tend to use other forms of alliances, e.g. spin-off.

\subsubsection{Venture capital}
Venture capital (VC) is a company mechanism to invest equity into independent firms directly or through venture capital fund (S33). Instead of having the process inside, the technology innovation is developed at that firm. There are two reasons why companies engage in VC: financial returns and strategic contribution to its business (S31). Through VC, company has access to latest knowledge and technologies. 

Study S27 finds that two or more companies can invest in the same independent startup and create a corporation syndication. In this context, the innovation produced by the firm can be accessed by all investors. This introduces a drawback because their fellow investor can use the outputs to harm the benefits of each corporate. Therefore, the study suggests either maximising isolationist (invest in many firms, but not as the central position) or minimising centralist (invest in few firms, but become the central position).

\subsubsection{Spin-Off}
Our study found that there is no agreement in literature on what the definition of spin-off is. The
term spin-off is also interchanged with internal venture, since their inceptions are similar. Both spin-offs and internal ventures are mainly used to facilitate a development of a product, which are new to the company. However, the whole resources used in the internal venture are coming from the parent company, which is not the case in the spin-offs \citep{roberts85,narayanan09}.

There are two main components in spin-off: technology from parent company and former employee \citep{carayannis98}. The technology which does not fit to the main business stream is transferred to the new company (S13,S14,S17,S28). It is initiated by parent company as the effect of spillovers of R\&D outputs (S13). On the other hand, spin-off is also initiated by former employee of a parent company who brings the knowledge or technology out \citep{carayannis98}. The decision to leave the company is typically because of the management rejection on the employee's ideas \citep{anton95}.

In both cases, the relatedness between parent company and the spin-off is higher at the early stage. This is due to the higher need of resources of the spin-off company (S14) and direct access to the market \citep{anton95}. Therefore, parent company usually still gets some shares as the compensation of the loss of resources and also as a mean to trace the direction of technology development (S14). Spin-off could be the solution to grow and nurture intrapreneurs (S11). For entrepreneurs, the ratio of risk/reward is higher than intrapreneurs, because it is shared among the shareholders.

Study S13 finds that spin-offs governed by outside investors show a higher financial performance than spin-offs led by insider CEO. The study also finds that spin-offs led by insider CEO limit their activities to the practices that are applied in the parent company. They are more likely to operate in the similar market as the parent company. In addition, spin-offs led by insider CEO lack of access to senior managers in other companies, which hinder them to recruit new senior manager to the venture.

\subsubsection{Crowdsourcing}
The term crowdsourcing was coined by \cite{howe06} to describe ``the phenomenon of outsourcing the tasks of the company by using the collective intelligence'' (S36). While outsourcing refers to a closed and certain entity, crowdsourcing is open and unlimited. By taking the advantage of Web 2.0, companies look for the suitable solutions from Internet users. Crowdsourcing is not always free. It reduces the cost of resources and manpower than in-house development as well as the communication cost. Crowdsourcing eliminates the costs for recruitment and hiring, training and supervising employees, as well as cost for creating a functional work environment in house (S36). Moreover, it gives an opportunity to users to influence the price. However, the participation of crowd in the innovation process is determined the company's reputation (S20).

Through crowdsourcing, companies obtain three benefits (S36). Firstly, product development through communities. This is participatory approach, where the communities are encouraged to give feedbacks on the existing products and proposed new ideas. Then, they are invited to vote for most interesting and promising ideas to be implemented. Second is solving existing problems. Through crowdsourcing, companies look for a solution beyond their competencies. It reduces the expense for human resources and development time. The third is innovation through crowd. In this approach, the users use the platform as the media to create new product and sell it with the obligation to pay royalties to the creator. Any user can come up with their innovative ideas and later on sell them through online stores.

In software industry, community is considered as a complementary asset (S19). Companies could collaborate with community (e.g. Free and Open Source Software community) to develop a new product or service. However, this is difficult to lock community to create value to one company only, since competitors may interfere. Moreover, since communities work without control and protection, companies need to deploy their employees in the communities to gain access to developments and convert the knowledge created in the community into a complementary asset.

The involvement of community is also recognised in social network service such as Facebook. Study S32 shows that Facebook open application programming interface (API) platform policy increases the involvement of the user in the innovative process. Rather than making it closed and develop everything in-house, Facebook invites third-parties to create applications on the top of the platform. This policy has significant impact on the growth of traffic data.

\subsection{Challenges of Different Types of Innovation initiative in Large Companies}
Table \ref{table:challenges} summarises the challenges of different types of innovation initiative in large companies, as described in Section \ref{sec:innovation_type}. Sources that are written in italics are studies in the context of large software companies. Our study identifies that in the free initiative has a specific challenge to both corporate management and the employees. For corporate management, free initiative requires a change in management style. Corporate management needs to provide new infrastructure that promote, encourage and nurture intrapreneurs e.g. policy, culture etc. In return, these changes could motivate employees to undertake entrepreneurial activities. 

Generating innovation is not always the main motive for companies to engage in alliances. Our study finds that companies involve in creating a joint venture is to open new market, and through crowdsourcing, companies are able to save costs for resources and manpower. In alliances, the different challenges faced by each initiative is related to joint activities with other parties, e.g. other companies, communities, etc.

\begin{table*}[htbp]
    \caption{Key challenges of each innovation initiative}
    \center
    \label{table:challenges}
    \begin{tabular}{|p{3cm}|p{7.4cm}|p{2.5cm}|}
    \hline 
    Innovation initiative & Identified challenges & Sources\\  \hline
    \multirow{2}{*}{Intrapreneurship} & For management: Change in management style, e.g. accepting failure as a learning process, balancing freedom to work in a different way and monitoring & S1, \textit{S2}, S4\\ \cline{2-3}
     & For employees: Gain trust from management & S21 \\ \hline
     \multirow{2}{*}{Bootlegging} & For management: Identify and manage the process & \textit{S8}, S16 \\ \cline{2-3}
     & For employees: Gaining support for top management to secure resources & \textit{S8}, S16 \\ \hline
    R\&D & Limited scope to the existing technology; balancing the time and effort to implement ideas from joint R\&D initiative & S11, S28 \\ \hline
    Internal venture & Setting up the new venture, e.g. members, process, resources, etc. & S12, S5, S25 \\ \hline
    Subsidiary & Cost to build specific product, lack of focus (too much entrepreneurship) and cost of administering the internal market & S9 \\ \hline
    Joint venture & Cannot be used to promote product innovation, but rather to expand the market & S30 \\ \hline
    Venture capital & Risk that investors could use the outputs to harm the benefits of each corporate &  S27 \\ \hline
    Spin-off & Setting up the new company, e.g. resources, members, technology, etc. & \textit{S13}, S14, S17, S28 \\ \hline 
    Crowdsourcing & Getting the participation of crowd and locking the crowd to create value to one company only & \textit{S19, S20} \\ \hline
    \end{tabular}
\end{table*}
\section{Interview Results Regarding Innovation in Large Software Companies}
\label{sec:initial_evaluation}

In this section, we present the results from the four interview respondents. The first part of the interview was to discuss different types of innovation initiative in large software companies. In their experience, all interviewees have seen how their companies conducted the innovation initiatives. Regarding the innovation initiative tree, one of the interviewees saw that it can be seen as the evolution of innovation initiatives in established companies. Companies could use certain initiatives to foster innovation in a specific condition:
\begin{displayquote}
\textit{I have seen free initiatives. I have seen bootlegging. I have seen intrapreneurship, expert systems where very typical during [a] certain time within [our company]. ... I saw in my experience those [initiatives] have been there in different times. That should be taken into account that certain methods have been active in different periods. For instance, joint venture and internal venture, they have, in my experience, clear time slot. They did not typically exist at the same time in the case of [our company]. ... It is a function of the evolution phase of the company.} -- INT\_3
\end{displayquote}

To their knowledge, all interviewees agree that the tree has captured all initiatives that has been done in their companies, except for intrapreneurship. One of the interviewees (INT\_4) explained that in his company, the free initiative is fostered through a mechanism called ``ground up innovation''. Unlike intrapreneurship as discussed in Section \ref{sec:intrapreneurship}, in this mechanism, employees have one working day per month that can be used within team to work on new idea in their own time. The idea could be shared in an internal website to get feedback from other employees. Then they have monthly demo day where they have two minutes of fame to present their ideas across R\&D offices. Unfortunately, even though some of the ideas are seemed to be a really good and viably, there is no clear mechanism to take those forward in the company. In addition, the introduction of new mechanism to promote free initiatives would raise the need a new mechanism to manage them:
\begin{displayquote}
\textit{If you want to utilise free initiatives, you need some kind of mechanism to handle them, evaluate them, boost them further, and that mechanism is additional to the normal manager's daily work.} -- INT\_3
\end{displayquote}

All interviewees agree that in many cases the motivation for companies to engage in those innovation initiatives is not because of innovation goal per se, but rather business as usual. It is typical for companies that are already close to the top of their market segment:
\begin{displayquote}
\textit{[Our] focus [in forming an alliance] was to somehow boost [our] solutions to the market. [It is] not utilised for gathering new technology or finding new innovations.} -- INT\_3
\end{displayquote}

\begin{displayquote}
\textit{The biggest acquisition we have done over the last six years was, we bought a cloud storage company in France. .. We were buying technology but eventually we realised that we bought customers because we had to rewrite the whole technology. But we were able to gain big customers we did not have in the past. Then we were able to upgrade their old technology with our [newest] technology [that] we had developed at the company. ... So we bought business. We thought we were buying technology but actually we bought business. } -- INT\_1
\end{displayquote}

In the second part of the interview, the interviewees were asked what they perceived as the key challenges for large software companies to engage in each type of innovation initiative. Table \ref{table:challenges_perception} summarises the key challenges of innovation initiative that have been perceived by industry practitioners.

\begin{table*}[htbp]
    \caption{Key challenges of innovation initiative perceived by practitioners}
    \center
    \label{table:challenges_perception}
    \begin{tabular}{|p{3.5cm}|p{8.4cm}|}
    \hline 
    Innovation initiative & Identified challenges \\  \hline
    Intrapreneurship & For management: to create culture and infrastructure that promotes and encourage innovation \\ \hline
    Internal venture & Balancing control and autonomy \\ \hline
    Joint venture & Can be used to generate innovation, but need to have consensus about the outputs among all collaborating companies  \\ \hline
    \end{tabular}
\end{table*}

One of the interviewees (INT\_2) argued that intrapreneurship in large software companies should be done in top-down fashion. It does not mean that the top management comes up with a specific idea to be implemented but allows employees to further investigate on ideas that fall into the company strategy. Otherwise, it will never get through:
\begin{displayquote}
\textit{If I create a new technology, it can be disruptive. It can be amazing but I need to convince so many layers on top of me. I need to get everybody moving in the same direction. First of all, top management needs to be in a ready to accept innovation in certain areas.} -- INT\_2
\end{displayquote}

As discussed in Section \ref{sec:internal_venture}, internal venture has been suggested a potential avenue for nurturing corporate innovation in large companies. It aims to develop new products or services which might be targeted at a new market. Thus, it allows companies to acquire new competence. However, as an organised initiative, internal venture is fully driven by company strategy. One of the interviewees (INT\_4) admitted that the challenge with internal venture is to balance between control and autonomy. 
\begin{displayquote}
\textit{Internal venture is quite big investment; a dedicated people, really taken away from whatever they were doing at that point. [The corporate management] decided when is about to stop, it is not the team that can decide that. [To get success], the top management has to understand that it is not about control. The more control you have, the less innovation you have. ... So top management needs to understand  that they have the money and should guard the property of the company but the evaluation of these ideas should be external.} -- INT\_4
\end{displayquote}

One of the interviewees (INT\_3) argued that joint venture can also be used to pursue innovation, in terms of intellectual property right (IPR). In a joint venture, all parent companies put a lot of effort e.g. knowledge, financial resources, manpower, to support innovation. However, the outcomes need to be managed properly to sustain the collaboration.
\begin{displayquote}
\textit{For instance [his former company] had a joint venture with [Company-X]. [Company-X] is currently leading the chipset market in the world for mobile phones. During late 90's and early 2000, [his former company] was cooperating in a joint venture with  [Company-X]. They ended up with a fight. The reason was some IPR issues, which is fairly not strong reasons to fight and they split.} -- INT\_3
\end{displayquote}
\section{Discussion}
\label{sec:discussion}

While a lot of research on corporate innovation has been done in different industries, our mapping study finds a small number of studies in the context of software industry. Table \ref{tab:innovation_context} shows that out of 38 primary studies, only 7 studies are conducted in the context of large software company. One of the reasons is in strategic level, large software companies share the same characteristics as in other large companies. They already have on going business and market that they must maintain (S26). Moreover, they also rely on bureaucracy and hierarchy to run the big and complex business \citep{ross87}. When it comes to innovation, large software companies deal with the similar issue as other companies. For instances, there are several layers of approval to get through to ensure its sustainability. They also need to ensure that any innovation initiatives would not harm their existing businesses \citep{ahmed98}. 

In tactical level, due to the similarity of its characteristics, the innovation initiatives in large software companies are also similar to other companies. Out of 7 studies in the context of large software company, the common innovation initiatives in large software companies are intrapreneurship, bootlegging, internal venture, spin off and crowdsourcing. Each initiative aims at radical innovation; something that has not been done in the company (S2, S8, S20, S23). Since it is the first time for companies, each initiative requires an autonomous environment for innovators to pursue and implement their ideas. 

In the operational level, our mapping study suggests that innovation in large software companies are influenced by the advancement of Internet technology. The introduction of Web 2.0 has shifted the focus of companies to provide product/service to individual users to community and emerged User Community Driven Internet (UCDI) ventures (S20,S23). To increase user growth, companies could take advantage of viral marketing strategy. 

Our mapping study results also suggest that innovation in large software companies is driven by open innovation paradigm. Open innovation paradigm advocates the use of external entities in the innovation process e.g. users, customers, community, etc. Early user integration in the front-end innovation is valuable for generating new ideas and identifying perceived value. Users are valuable as an asset to large companies because their needs are the source of innovative ideas \citep{edison13}. When it comes to product or service development, our reviewed studies reveal that companies engage with FOSS movement to gain access to the technology developed in the community, which then can be used in the internal development (S19). Another way to increase the participation from the crowd is through Open API, which allows third party to create their own APIs and facilitate group interactions. Facebook, Google, Apple, to name a few, are example of large software companies that have publicly opened their APIs.

Our interview results shows that one of the motives why large software companies involved in such initiative is business as usual. Large software companies do not lack creative people (S2). However, they are bounded to work only on the company strategy. When they want to enter into an going-competition which is new for them, they have two options. First, they could buy a new emerging company with innovative and promising product, and merge them into their structure. In this way, they could save few years to develop the technology from scratch. Second, they could join an alliance as a strategy to increase market share. 

Another finding from our interview results is that, often free initiative in large companies is limited only to a current product in the market, as part of main product line development activities. The initiative is not intended as a vehicle to develop a new product and bring it into the market. It requires new mechanisms to handle and push the initiative forward.

A recent study in the context of large software companies by \cite{jarvinen14} introduces the concept of Mercury business, which is inspired by the Lean startup principle \citep{ries11}. In Mercury business, companies not only seek for a business opportunity but also experiment and pivot existing business to a new area. However, Mercury business is aimed at improving the current product or services. Lean startup principle is also adopted in R\&D based product innovation. \cite{bosch12} proposes an innovation experimentation system to minimise R\&D investment and increase customer satisfaction. However, this method is limited to SaaS and embedded system.

Our mapping study results also identify the key challenges of each innovation initiative in the context of large software companies. While literature has suggested the importance of free initiative (e.g. S1,S2,S3,S4,S10,S12,S15,S18,S21,S37,S38), it needs a change of management style. This is also exemplified by one of our interviewees (INT\_3), who argue that free initiative would introduce a new task for managers.

Unlike the findings from our mapping study, one of the interviewees explained that Joint Venture is not only for expanding the market but can also be used to generate innovation, e.g. to generate IPR. However, the IPR is owned by Joint Venture rather than a single company, therefore it can be used for any development activities in each collaborating company.
\section{Conclusion and Future Work}
\label{sec:conclusion}

While entrepreneurship emphasises on exploring and exploiting new opportunities to create wealth, innovation is a specific tool to find those opportunities. This paper presents an in-depth review of innovation initiative in large companies. A total of 38 primary studies were identified which discussed how large companies continuously innovate. As a result 9 different types of innovation initiative were found.

Our study results find that in strategic and tactical level, there is no difference between large software companies and other large companies, whilst in the operational level, large software companies are influenced by Internet technology. In addition, large software companies use open innovation paradigm as part of their innovation initiatives. Our interview results confirm that all initiatives are also practised in large software companies. However, not all of them are implemented as a mean to generate innovation, but rather business as usual.

Our mapping study results also suggest that there is a lack of high quality empirical studies on innovation initiative in the context of large software companies. A total of 7 studies are conducted in the context of large software companies, which reported 5 types of innovation initiatives: intrapreneurship, bootlegging, internal venture, spin-off and crowdsourcing. 

In terms of theoretical implication, the paper makes three contributions to both software engineering research and practice. First, this paper represents the overview of the current research on innovation initiatives inside large software companies. Research could use our results to navigate their research focus to address the identified gap. The second contribution of this paper is to provide an innovation initiative tree. The tree shows that innovation in large companies can be initiated bottom-up or top down, either by individual employee or management. Moreover, the tree distinguishes the initiatives that happen inside and outside the company. This paper also identifies which innovation initiatives are found in the context of large software companies.

The third contribution is for practitioners; our study also identifies the key challenges faced by each innovation initiatives found in the literature. The study is extended with interviews with four industry practitioners with vast years of experience in innovation initiative in the context of large software companies. Industry practitioners could use our findings to reflect on their experience on corporate innovation in order to minimise the challenges in their context.

We envision our future work is to further empirically evaluate the innovation initiative tree in large software companies. To improve the generalisability of our findings, more practitioners and researchers from large software companies should be involved in the future studies. Another future work could look at specifically on product innovation. As described in Section \ref{sec:discussion}, there is an increasing interest to adopt Lean startup principle to facilitate large software companies to radically generate product innovation. 

In addition, it is possible that adding non-software and information system related databases may yield similar or different findings. The comparison of findings from a different databases and the findings presented in this study can potentially be considered as future work.

\section*{References}
\bibliographystyle{apalike}
\bibliography{elsarticle}
\end{document}